\documentclass[twocolumn,showpacs,preprintnumbers,amsmath,amssymb, prb, reprint]{revtex4-1}

\usepackage{stmaryrd}
\usepackage{txfonts}
\usepackage{amssymb}
\usepackage{mathrsfs}
\usepackage{graphicx}
\usepackage{dcolumn}
\usepackage{bm}
\usepackage{epsfig}
\usepackage{color}                    
\usepackage{hyperref}                 

\begin{document}

\preprint{\href{http://dx.doi.org/10.1103/PhysRevB.88.060404}{S.-Z. Lin and L. N. Bulaevskii, Phys. Rev. B {\bf 88}, 060404(R) (2013).}}

\title{Quantum motion and level quantization of a skyrmion in a pinning potential in chiral magnets}

\author{Shi-Zeng Lin}
\affiliation{Theoretical Division, Los Alamos National Laboratory, Los Alamos, New Mexico 87545, USA}

\author{Lev N. Bulaevskii}
\affiliation{Theoretical Division, Los Alamos National Laboratory, Los Alamos, New Mexico 87545, USA}

\begin{abstract}
A new topological excitation called skyrmion has been observed experimentally in chiral magnets without spatial inversion symmetry. The dynamics of a skyrmion is equivalent to an electron moving in a strong magnetic field. As a skyrmion involves large number of spins, it is not clear whether there exist observable quantum effects. In this work, we study the quantum dynamics of a skyrmion in a pinning potential. Without a pinning potential, the skyrmion performs cyclotron motion due to the strong emergent magnetic field originating from the Berry phase of spins, and all skyrmions occupy the lowest Landau level. Their wave functions are strongly localized in a region with radius less than $1\ \AA$ when no driving force is applied. Thus in most circumstances, the quantum effects of a skyrmion are weak. In the presence of a pinning potential, the lowest Landau level for skyrmions is split into quantized levels, characterized by the orbital moments. The transition between quantized levels can be observed experimentally by microwave absorption measurements in low temperature region. The quantum effects are more prominent for a skyrmion with a small size, which can be realized in magnets with a large Dzyaloshinskii-Moriya interaction.
\end{abstract}
\date{\today}
\maketitle

Recently stable triangular lattice of skyrmions in chiral magnets without inversion symmetry has been observed in conductors, \cite{Muhlbauer2009,Munzer10,Pfleiderer10,Yu2010a,Yu2011,Heinze2011} such as MnSi, $\rm{Fe_{0.5}Co_{0.5}Si}$ and also in insulator, \cite{Seki2012} $\rm{Cu_2OSeO_3}$, by the neutron scattering, Lorentz transmission electron microscope and spin-polarized scanning tunneling microscopy. In the temperature-magnetic field phase diagram, there are magnetic spiral phase, skyrmion lattice, ferromagnetic and paramagnetic states.  In bulk samples, the skyrmion phase occupies a small portion of the phase diagram, \cite{Muhlbauer2009} while in thin films skyrmion lattice occupies a large area in the phase diagram. \cite{Yu2011} For their promising applications in spintronics, skyrmions have attracted considerable attention since the experimental observations. 

\begin{figure}[t]
\psfig{figure=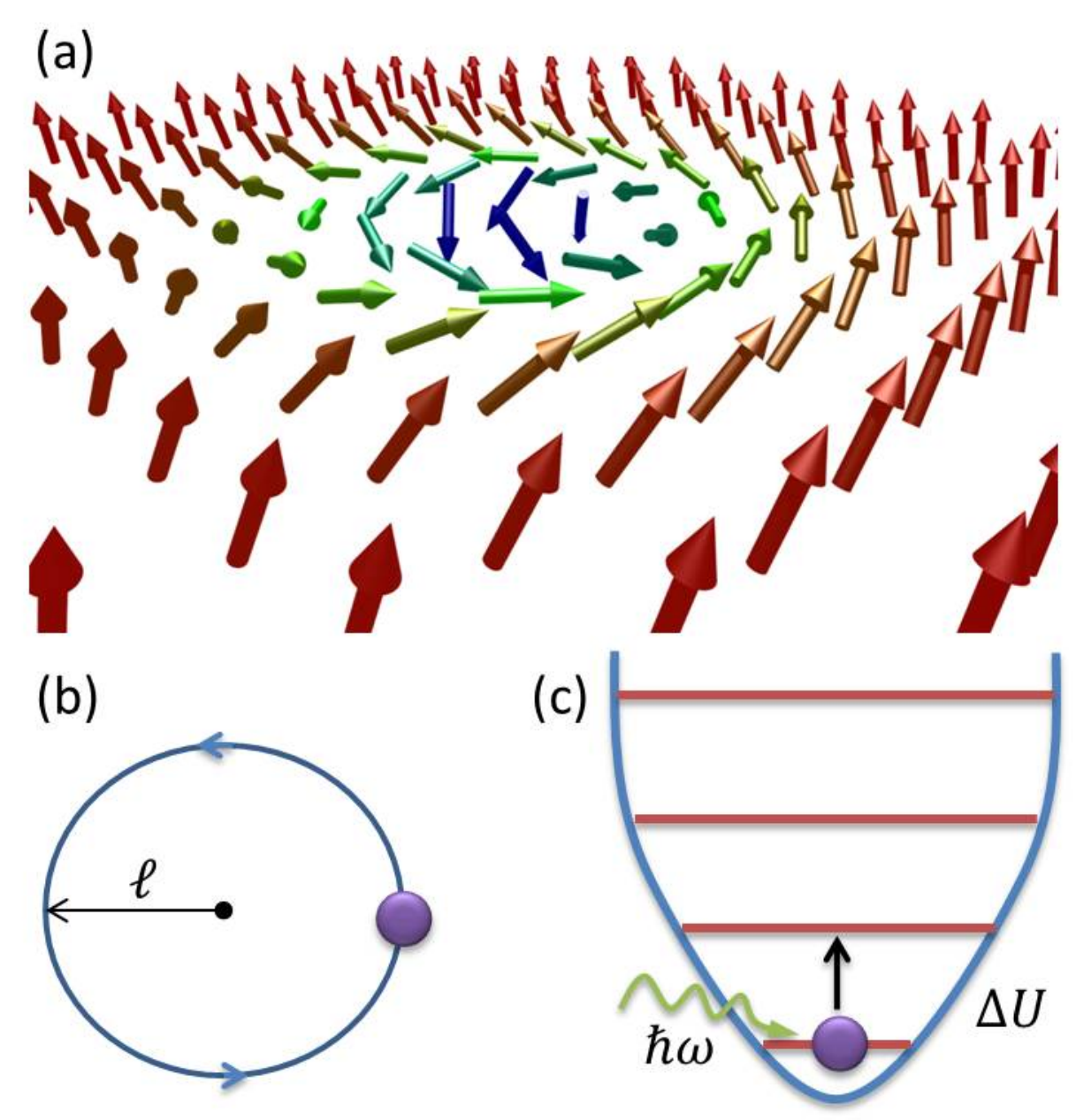,width=\columnwidth}
\caption{\label{f1}(color online) (a) Schematic view of the spin texture in a skyrmion. (b) The dynamics of skyrmions is equivalent to electrons moving in a strong magnetic field. The skyrmion performs cyclotron motion and its wave function is strongly localized in the absence of an external drive. (c) In the presence of a pinning potential, the LLL of skyrmion is split into discrete levels. Under a microwave irradiation, the skyrmion can be excited to a high energy level. }
\end{figure}

A skyrmion is a mesoscopic topological excitation in the ferromagnetic ground state. In skyrmion the spins wrap a sphere when one goes from the center of the skyrmion to the infinity, see Fig. \ref{f1} (a). When an external magnetic field is applied, skyrmions can be stabilized either by the Dzyaloshinskii-Moriya (DM) interaction \cite{Bogdanov89,Bogdanov94,Rosler2006} or by the dipole-dipole interaction \cite{Yu2012b,Finazzi2013}. Due to the topological nature, the skyrmions inside samples are stable against weak perturbations and disorder.  Typically the size of skyrmion is about $10$ to $100$ nm and involves $10^3$ to $10^5$ spins.  In the conducting compounds where the Hund coupling is much stronger than the Fermi energy, the skyrmion can be driven by a spin polarized electric current through the spin transfer torque \cite{Bazaliy98,Li04,Tatara2008}, i.e. transferring of the conduction electron's magnetic moments to localized spin moments; For the insulator $\rm{Cu_2OSeO_3}$ with a strong magnetoelectric coupling, \cite{Seki2012b} it was proposed to drive the skyrmion by an electric field gradient \cite{Liu2013b} because skyrmion carries electric polarization. It was shown experimentally that the pining of skyrmion by quenched disorders is extremely weak. \cite{Jonietz2010,Yu2012,Schulz2012} The depinning current density for skyrmions is found to be about $10^5 \sim 10^6\ \rm{A/m^2}$, four order of magnitude smaller than that for the magnetic domain walls. 

Since a skyrmion involves huge amount of spin degrees of freedom, at the first sight one might expect that the quantum effects are weak. To date, most theoretical work on skyrmion treats it as a classical topological excitation. If these spins change collectively, observable quantum behavior of skyrmion may arise under certain conditions, and it is not clear how to observe the quantum effects in experiments. In the present work, we will show that the quantum motion of skyrmion in the presence of defects can result in energy level quantization. The quantized levels can be observed experimentally by microwave absorption or by an ac driving force. 

For a skyrmion, besides the extended modes (continuum of the spin wave spectrum), additional localized modes inside the skyrmion (internal modes) have been identified. \cite{szlin13skyrmion4} One is the Goldstone mode corresponding to the translation motion of skyrmion. The other internal modes, such as the breathing mode, are gapped. If the internal modes are not excited, we can consider only the translational motion of skyrmion and treat skyrmion as a rigid particle. We consider the two dimensional motion of skyrmions in a thin film. In this case, the equation of motion for skyrmions is governed by \cite{szlin13skyrmion2}
\begin{equation}\label{eq1}
\eta\mathbf{v}_i=\mathbf{F}_M+\sum_j \mathbf{F}_{ss}(\mathbf{r}_j-\mathbf{r}_i)+\sum_j\mathbf{F}_p(\mathbf{r}_j-\mathbf{r}_i)+\mathbf{F}_d,
\end{equation}
where $\mathbf{v}_i=(v_{ix},\ v_{iy})$ is the skyrmion velocity. The term on the left-hand side is the dissipative force with a damping coefficient $\eta$. It accounts for the damping of skyrmion motion and is induced by the underlying damping of the spin relaxation. The mass is zero for a rigid skyrmion. When the internal modes are excited, for instance when a skyrmion passes through a defect \cite{Iwasaki2013}, there will be additional contributions to the damping and inertial term \cite{Makhfudz2012,Mochizuki2012}, similar to vortices in superconductors \cite{Blatter94}. In this case, we need to add a term $m_s \partial_t\mathbf{v}_i$ to the left-hand side of Eq. \eqref{eq1} with $m_s$ being the skyrmion mass. The first term at the right-hand side of Eq. \eqref{eq1} $\mathbf{F}_M=4\pi\gamma^{-1}\hat{z}\times \mathbf{v}_i$ is the Magnus force per unit length, which is perpendicular to the velocity.  Here $\hat{z}$ is a unit vector normal to the film, and $\gamma=a^3/(\hbar S)$ with $a$ the lattice constant of magnetic unit cell of chiral magnets and $S$ the spin of the localized moments. Further $\mathbf{F}_{ss}$ is the pairwise interaction between two skyrmions at $\mathbf{r}_i$ and $\mathbf{r}_j$. The term $\mathbf{F}_p$ is the interaction between a skyrmion at $\mathbf{r}_i$ and a quenched disorder at $\mathbf{r}_j$. The last term $\mathbf{F}_d$ is the driving force.

In conductors the driving force is due to the spin-transfer torque exerted by a spin polarized electric current $\mathbf{J}$. It is given by $\mathbf{F}_d=\hat{z}\times \mathbf{J}\Phi_0/c$ which is nothing but the Lorentz force, arising from the emergent quantized magnetic flux $\Phi_0=hc/e$ carried by the skyrmion in the presence of a finite current. For insulators, the force can be produced by the gradient of electric field $\mathbf{E}$ and it is given by ${F}_{d,\mu}=\mathbf{P}\cdot\partial_\mu \mathbf{E}$, where $\mathbf{P}$ is the total electric polarization per unit length inside a skyrmion and $\mu=x,\ y$. \cite{Liu2013b} Equation \eqref{eq1} also describes the dynamics of vortices in type II superconductors in the presence of magnetic fields.   \cite{Blatter94,Kopnin1995,Bulaevskii1995} For vortices, the dissipative force is dominant over the Magnus force in most parameter space; while for skyrmions, the dissipation is due to the damping of precession of spins and is usually weak, thus the Magnus force is dominant.

In real films the defects in the crystal structure cause pinning of skyrmions. Pinning potential can also be introduced intentionally with its strength and size being controllable in experiments (for example creation of columnar defects). For a properly engineered pinning potential, the cyclotron motion of skyrmion in the pinning potential in the quantum region produces experimentally observable consequences.

Equation \eqref{eq1} also describes an electron moving in an effective magnetic field $H^*=2\Phi_0 d/a^3$ for $S=1$. For film thickness of $d=10$ nm, the field is $H^*=10^6$ T for $a\approx 3\ \AA$. With such a huge emergent internal field $H^*$, skyrmions are strongly confined in their cyclotron orbital with radius $\ell=a\sqrt{a/(4\pi d)}\approx 0.15\ \AA$. Although $\ell$ is much smaller than the lattice constant, $\ell\ll a$, we can still use the continuum motion of skyrmion in Eq. \eqref{eq1} for the following reason. The position $\mathbf{r}$ in Eq. \eqref{eq1} is the skyrmion center of mass defined as $\mathbf{r}=\int dr'^2\mathbf{r}'Q(\mathbf{r}')/\int dr'^2Q(\mathbf{r}')$, where $Q(\mathbf{r}')={\bf{n}}\cdot({\partial _x}{\bf{n}} \times {\partial _y}{\bf{n}})/(4\pi)$ is the topological charge density for skyrmion and $\int dr'^2Q(\mathbf{r}')=\pm 1$ for skyrmion and anti-skyrmion. Here $\mathbf{n}$ is a unit vector representing the direction of the spin at position $\mathbf{r'}$. Since the size of skyrmion, typically 10 nm to 100 nm, is much larger than the lattice constant, the skyrmion center of mass $\mathbf{r}$ is a continuous variable, which justifies the continuum motion of skyrmion in Eq. \eqref{eq1}.

The skyrmion density can be controlled by applied magnetic fields. Near the phase boundary between the ferromagnetic phase and the skyrmion lattice phase, the density of skyrmion is small. In these cases, we can neglect the interaction between skyrmions putting $F_{ss}=0$. Neglecting the weak dissipation and using the canonical quantization, the Hamiltonian corresponding to Eq. \eqref{eq1} with an inertia term, $m_s \partial_t\mathbf{v}_i$, is 
\begin{equation}\label{eq2}
\mathcal{H}(\mathbf{r})=\frac{1}{2}\left[(-i\partial_x+A_x)^2+ (-i\partial_y+A_y)^2\right]+U_p(\mathbf{r})-F_d x,
\end{equation}
where $\mathbf{r}=(x,\ y)$, $\mathbf{A}$ is the vector potential and $U_p$ is the pinning potential. The driving force is assumed along the $x$ direction. Here dimensionless units are used. Energy is in units of $\hbar\omega_c=4\pi\hbar^2 d/(a^3 m_s)$; length is in units of $\ell$; magnetic field is in units of $H^{*}$ and force is in units of $\hbar\omega_c/\ell$. We note that the skyrmion mass cannot be zero in the quantum region, otherwise the energy of zero-point motion of skyrmion diverges. This means that the quantum motion can also distort skyrmions and excite internal modes, which contribute to the skyrmion mass. However we can still use the particle description with a mass term for skyrmions in Eq. \eqref{eq1}, since a skyrmion is a well-defined topological excitation. Since the mass of skyrmion is due to the excitation of internal modes, we expect the mass of the skyrmion is small.  In our calculations we will take $m_s\rightarrow 0$ limit at the final step. Without driving force, $F_d=0$, and at pinning potential $U_p=0$, the energy levels for the Hamiltonian in Eq. \eqref{eq2} are the well-known Landau levels with spacing $\hbar\omega_c$. In the limit $m_s\rightarrow 0$, it is sufficient to consider the lowest Landau level (LLL), where the skyrmion wave function under the symmetric gauge $\mathbf{A}=(y,\ -x)/2$ is
\begin{equation}\label{eq3}
\Psi_n(\theta, \mathbf{r})=\frac{r^n}{\sqrt{\pi n!}}\exp(i n \theta-r^2/2),
\end{equation}
with an integer $n$ corresponding to the orbital momentum along the $z$ axis. As shown in Fig. \ref{f1} (b), the skyrmion wave function is strongly localized in the region with a radius $\ell$, which is just the zero-point motion of skyrmion. As $\ell$ is much smaller than the skyrmion size, the quantum effects of skyrmion in most cases are weak. The density of states in the LLL is given by $N=H^*\mathcal{A}/\Phi_0=2N_s$ with $\mathcal{A}$ being the lateral area of the film and $N_s$ the total number of spins, which is far more bigger than the number of skyrmions. Thus it is impossible to fill the LLL by skyrmions, and the integer quantum Hall effect for skyrmions is absent.  

When a driving force is applied, $F_d>0$, skyrmion wave function is delocalized and is given by
\begin{equation}
\Psi_{k_y}(\mathbf{r})=\frac{1}{\sqrt{\sqrt{\pi} L_y}}\exp\left[i k_yy+\frac{i x y}{2}-\frac{(x+k_y-F_d)^2}{2} \right],
\end{equation}
where $k_y=2m\pi/L_y$ with $L_y$ being the length along the $y$ direction and $m$ an integer. These states are delocalized along the strip with the center at $x=F_d-k_y$ with width $\ell$. The corresponding energies are $E_{k_y}=k_y F_d-F_d^2/2$.

\begin{figure}[t]
\psfig{figure=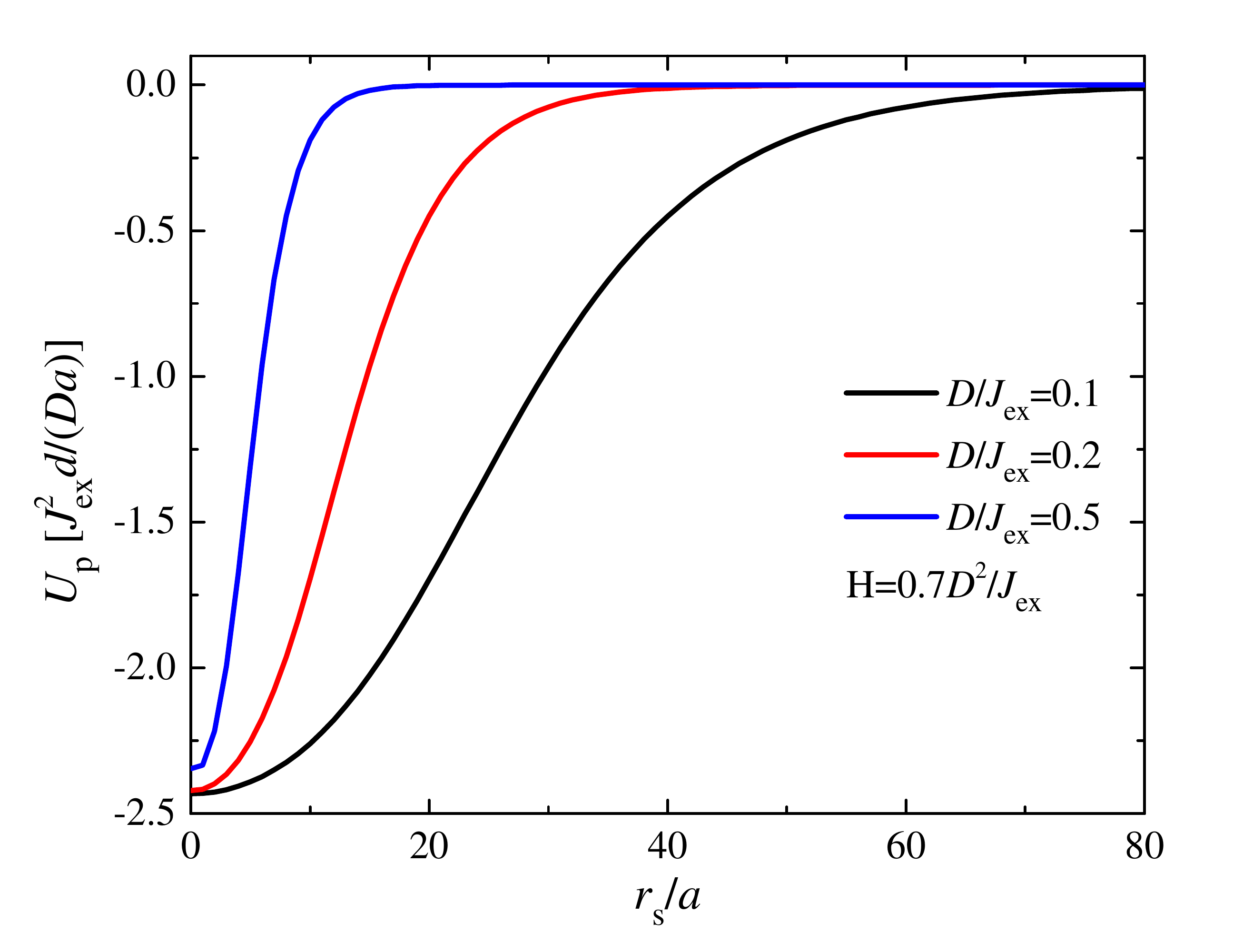,width=\columnwidth}
\caption{\label{f2}(color online) Pinning potential $U_p$ versus the separation $r_s$ between the center of the defect and the center of the skyrmion. The center of the defect is at $\mathbf{r}=0$ and the size of the defect is $\xi_d$. The defect is modeled by  $J_{\rm{ex}}'(\mathbf{r})=[J_{\rm{ex}}-J_d \exp(-r/\xi_d)]$ with $J_d=0.8 J_{\rm{ex}}$ and $\xi_d/a=J_{\rm{ex}}/D$. The external field is $H=0.7D^2/J_{\rm{ex}}$.}
\end{figure}

In the presence of the pinning potential $U_p$ and without a driving force, $F_d=0$, the degeneracy of the LLL with different orbital moment $n$ is lifted, see Fig. \ref{f1} (c). The energy for the orbital moment $n$ is
\begin{equation}\label{eq4}
E_n=\frac{2}{n!}\int_0^{\infty}dr U_p(r) r^{2n+1}\exp(-r^2).
\end{equation}
For convenience of calculations, we assume the pinning potential has the form
\begin{equation}\label{eq5}
U_p(r)=-U_0 \frac{r_p^2}{r^2+r_p^2},
\end{equation}
where $U_0$ is the pinning strength and $r_p$ is the size of the pinning potential. For columnar pinning potential with the size of order of the skyrmion size as shown below (see Fig. \ref{f2}), $U_0\sim J_{\rm{ex}}^2 d/(Da)$ and $r_p/a\sim J_{\rm{ex}}/D$, where $J_{\rm{ex}}$ is the exchange interaction and $D$ is the DM interaction. \cite{Zang11} Note that the columnar pins can be created by heavy ions bombardment similar to that for superconductors. \cite{Civale1991,Konczykowski1991} The quantum level for a skyrmion in this pinning potential becomes equidistant and is given by
\begin{equation}\label{eq6}
\Delta U=E_{n+1}-E_n\approx 2 U_0(\ell/r_p)^2.
\end{equation}
For the pinning potential Eq. \eqref{eq5}, we estimate $\Delta U\approx D/(2\pi)$. For MnSi, we have $J_{\rm{ex}}\approx 3$ meV and $D\approx 0.3$ meV \cite{Zang11}, thus $\Delta U\approx 0.05$ meV. The quantized levels can be observed when the temperature is lower than $T\ll 0.5$ K. For a stronger DM interaction $\Delta U$ becomes larger, which is helpful for experimental observations.

Here we discuss the transition of skyrmion between different discrete levels in the pinning potential. The skyrmion can be excited to the higher level by an ac driving force $F=F_{\rm{ac}}\cos(\omega t)$. When the frequency matches the level spacing $\hbar\omega=\Delta U$, there will be resonant excitation of skyrmion to the high levels. The transition probability is determined by the matrix element $\Gamma_{n, n+1}$,
\begin{equation}\label{eq7}
\Gamma_{n, n+1}\equiv F_{\rm{ac}}\langle n|x|n+1\rangle=\frac{F_{\rm{ac}}\sqrt{\pi}\ell (2n+1)!!}{2^{n+2}\sqrt{n!(n+1)!}}.
\end{equation}
For $\Delta U\approx 0.05$ meV, the corresponding angular frequency is about $\omega\approx 100$ GHz. This frequency is higher than those of internal modes for skyrmions \cite{Mochizuki2012}, indicating that skyrmions are distorted and internal modes are excited under irradiation of microwave at this frequency. Experimentally the transition between different quantum levels of skyrmion can be excited by applying microwave irradiation. For insulators like $\rm{Cu_2OSeO_3}$, the microwave interacts strongly with the skyrmion. While for conductors like MnSi, the film thickness should be much less than the skin depth. For thickness of tens of nanometers, the microwave can fully penetrate through the film and provides strong coupling with skyrmions.  

Skyrmions need to be confined in the pinning potential during the measurements. In reality, skyrmions can escape the pinning potential via thermal and/or quantum fluctuations. The thermally-assisted escaping of the skyrmion can be minimized by reducing the temperature well below the pinning strength, i.e. $k_B T\ll U_0$, with $k_B$ the Boltzmann constant. The quantum tunneling rate of skyrmion from one pinning center to its nearest pinning center with a separation $q$ is $\exp(-S_q)$ with $S_q\sim  q^2 d /a^3$. \cite{Feigelman1993,szlin13skyrmion2} Thus the quantum tunneling rate of skyrmions is negligible provided $q\gg a$.   

For a system with randomly distributed pining centers with varying pinning strength $U_0$ and pinning size $r_p$, $\Delta U$ may change from pinning centers to pinning centers. On average in the sample, the quantization levels become quasicontinuum, which poses a grand challenge for experimental observations. To overcome this difficulty, one may use a nanodisk of chiral magnets, where only one skyrmion can be stabilized in this confined geometry. \cite{szlin13skyrmion3} The pinning center can be introduced through heavy ion bombardment at the center of the disk. 

Finally we discuss the interaction between a skyrmion and a defect. We note a microscopic modeling of the defect has not been achieved yet. In the presence of defects in the crystal structure, both $J_{\rm{ex}}$ and $D$ becomes spatially dependent. Moreover the magnetostriction will also present, which increases the effective size of defects. To model the interaction between skyrmions and defects, one has to consider these effects, which is beyond the scope of the present study.  As a simple estimate sufficient for our purpose, we take a phenomenological approach to model the defect by a spatially dependent exchange interaction. The interaction between spins is given by the Hamiltonian\cite{Bogdanov89,Bogdanov94,Rosler2006,Han10,Rossler2011}  
\begin{equation}\label{eq9}
{\cal H}_s = -\frac{d}{a}\sum_{\langle i, j\rangle}\left({J_{\rm{ex}}'(\mathbf{r})}{{\bf{n}}_{i}}\cdot{{\bf{n}}_{j}} + {{\bf{D}}_{ij}}\cdot({{\bf{n}}_i} \times {{\bf{n}}_{j}})\right) -\frac{d}{a}\sum_i{\bf{H}}\cdot{{\bf{n}}_i}
\end{equation}
where summation is over the nearest neighbors $\langle i,\ j\rangle$, $\mathbf{D}_{ij}=D \hat{e}_{ij}$ is the in-plane DM interaction and spins are represented by a unit vector $\mathbf{n}$. Here $\hat{e}_{ij}$ is the unit vector from the site $i$ to the site $j$. Without defects $J_{\rm{ex}}'(\mathbf{r})=J_{\rm{ex}}$, the skyrmion phase is thermodynamically stable in the region $0.2 D^2/J_{\rm{ex}}<H<0.8 D^2/J_{\rm{ex}}$.  \cite{Han10,Rossler2011}  We model the defect by taking $J_{\rm{ex}}'(\mathbf{r})=[J_{\rm{ex}}-J_d \exp(-r/\xi_d)]$ with $J_d$ and $\xi$ being the strength and size of the defect respectively. Note that even for an atomic sized defect $\xi_d\approx a$, the pinning potential extends over the whole size of the skyrmion. To give an estimate, we assume that the distortion of the skyrmion induced by the defect is weak. We first obtain the solution of a skyrmion without the defect. We then substitute the obtained $\mathbf{n}_i$ into Eq. \ref{eq9} and calculate the self-energy of the skyrmion $E_s$ by fixing the separation $r_s$ between the center of the skyrmion and the defect. \cite{szlin13skyrmion2} The difference between the self-energy of a skyrmion in the presence of a defect and self-energy of a skyrmion without a defect is just the pinning energy $U_p=E_s$. The results of $U_p$ as a function of $r_s$ are shown in Fig. \ref{f2}. From Fig. \ref{f2}, we can see the strength of the pinning is about $J_{\rm{ex}}^2 d/(D a)$, which is a fraction of the self-energy of the skyrmion in the absence of a defect. The size of the pinning potential is $r_p/a\approx J_{\rm{ex}}/D$, which roughly is just the radius of skyrmion. For a larger $D$, the radius of the skyrmion is smaller thus the quantum effects become more pronounced.

To summarize, we have studied the quantum motion of a skyrmion in a pinning potential. The zero point motion of skyrmion is the cyclotron motion with radius less than $1\ \AA$. The skyrmions occupy the lowest Landau level with their wave functions strongly localized in the absence of a driving force. Thus in most circumstances, the quantum effect of a skyrmion is weak. In the presence of a pinning potential, the degenerate lowest Landau level split into discrete levels. We have shown that for a properly engineered pinning potential, these discrete levels can be observed experimentally through microwave absorption measurements at low temperatures. The quantum effects for skyrmions are more evident with a small skyrmion size, which can be achieved in magnets with a strong Dzyaloshinskii-Moriya interaction.

\begin{acknowledgements}
The authors are grateful for the helpful discussions with Cristian D. Batista, Avadh Saxena and Charles Reichhardt. This publication was made possible by funding from the Los Alamos Laboratory Directed Research and Development Program, project number 20110138ER.
\end{acknowledgements}


\begin{thebibliography}{35}%
\makeatletter
\providecommand \@ifxundefined [1]{%
 \@ifx{#1\undefined}
}%
\providecommand \@ifnum [1]{%
 \ifnum #1\expandafter \@firstoftwo
 \else \expandafter \@secondoftwo
 \fi
}%
\providecommand \@ifx [1]{%
 \ifx #1\expandafter \@firstoftwo
 \else \expandafter \@secondoftwo
 \fi
}%
\providecommand \natexlab [1]{#1}%
\providecommand \enquote  [1]{``#1''}%
\providecommand \bibnamefont  [1]{#1}%
\providecommand \bibfnamefont [1]{#1}%
\providecommand \citenamefont [1]{#1}%
\providecommand \href@noop [0]{\@secondoftwo}%
\providecommand \href [0]{\begingroup \@sanitize@url \@href}%
\providecommand \@href[1]{\@@startlink{#1}\@@href}%
\providecommand \@@href[1]{\endgroup#1\@@endlink}%
\providecommand \@sanitize@url [0]{\catcode `\\12\catcode `\$12\catcode
  `\&12\catcode `\#12\catcode `\^12\catcode `\_12\catcode `\%12\relax}%
\providecommand \@@startlink[1]{}%
\providecommand \@@endlink[0]{}%
\providecommand \url  [0]{\begingroup\@sanitize@url \@url }%
\providecommand \@url [1]{\endgroup\@href {#1}{\urlprefix }}%
\providecommand \urlprefix  [0]{URL }%
\providecommand \Eprint [0]{\href }%
\providecommand \doibase [0]{http://dx.doi.org/}%
\providecommand \selectlanguage [0]{\@gobble}%
\providecommand \bibinfo  [0]{\@secondoftwo}%
\providecommand \bibfield  [0]{\@secondoftwo}%
\providecommand \translation [1]{[#1]}%
\providecommand \BibitemOpen [0]{}%
\providecommand \bibitemStop [0]{}%
\providecommand \bibitemNoStop [0]{.\EOS\space}%
\providecommand \EOS [0]{\spacefactor3000\relax}%
\providecommand \BibitemShut  [1]{\csname bibitem#1\endcsname}%
\let\auto@bib@innerbib\@empty
\bibitem [{\citenamefont {M\"{u}hlbauer}\ \emph {et~al.}(2009)\citenamefont
  {M\"{u}hlbauer}, \citenamefont {Binz}, \citenamefont {Jonietz}, \citenamefont
  {Pfleiderer}, \citenamefont {Rosch}, \citenamefont {Neubauer}, \citenamefont
  {Georgii},\ and\ \citenamefont {B\"{o}ni}}]{Muhlbauer2009}%
  \BibitemOpen
  \bibfield  {author} {\bibinfo {author} {\bibfnamefont {S.}~\bibnamefont
  {M\"{u}hlbauer}}, \bibinfo {author} {\bibfnamefont {B.}~\bibnamefont {Binz}},
  \bibinfo {author} {\bibfnamefont {F.}~\bibnamefont {Jonietz}}, \bibinfo
  {author} {\bibfnamefont {C.}~\bibnamefont {Pfleiderer}}, \bibinfo {author}
  {\bibfnamefont {A.}~\bibnamefont {Rosch}}, \bibinfo {author} {\bibfnamefont
  {A.}~\bibnamefont {Neubauer}}, \bibinfo {author} {\bibfnamefont
  {R.}~\bibnamefont {Georgii}}, \ and\ \bibinfo {author} {\bibfnamefont
  {P.}~\bibnamefont {B\"{o}ni}},\ }\href {\doibase 10.1126/science.1166767}
  {\bibfield  {journal} {\bibinfo  {journal} {Science}\ }\textbf {\bibinfo
  {volume} {323}},\ \bibinfo {pages} {915} (\bibinfo {year}
  {2009})}\BibitemShut {NoStop}%
\bibitem [{\citenamefont {M\"unzer}\ \emph {et~al.}(2010)\citenamefont
  {M\"unzer}, \citenamefont {Neubauer}, \citenamefont {Adams}, \citenamefont
  {M\"uhlbauer}, \citenamefont {Franz}, \citenamefont {Jonietz}, \citenamefont
  {Georgii}, \citenamefont {B\"oni}, \citenamefont {Pedersen}, \citenamefont
  {Schmidt}, \citenamefont {Rosch},\ and\ \citenamefont
  {Pfleiderer}}]{Munzer10}%
  \BibitemOpen
  \bibfield  {author} {\bibinfo {author} {\bibfnamefont {W.}~\bibnamefont
  {M\"unzer}}, \bibinfo {author} {\bibfnamefont {A.}~\bibnamefont {Neubauer}},
  \bibinfo {author} {\bibfnamefont {T.}~\bibnamefont {Adams}}, \bibinfo
  {author} {\bibfnamefont {S.}~\bibnamefont {M\"uhlbauer}}, \bibinfo {author}
  {\bibfnamefont {C.}~\bibnamefont {Franz}}, \bibinfo {author} {\bibfnamefont
  {F.}~\bibnamefont {Jonietz}}, \bibinfo {author} {\bibfnamefont
  {R.}~\bibnamefont {Georgii}}, \bibinfo {author} {\bibfnamefont
  {P.}~\bibnamefont {B\"oni}}, \bibinfo {author} {\bibfnamefont
  {B.}~\bibnamefont {Pedersen}}, \bibinfo {author} {\bibfnamefont
  {M.}~\bibnamefont {Schmidt}}, \bibinfo {author} {\bibfnamefont
  {A.}~\bibnamefont {Rosch}}, \ and\ \bibinfo {author} {\bibfnamefont
  {C.}~\bibnamefont {Pfleiderer}},\ }\href {\doibase
  10.1103/PhysRevB.81.041203} {\bibfield  {journal} {\bibinfo  {journal} {Phys.
  Rev. B}\ }\textbf {\bibinfo {volume} {81}},\ \bibinfo {pages} {041203}
  (\bibinfo {year} {2010})}\BibitemShut {NoStop}%
\bibitem [{\citenamefont {Pfleiderer}\ \emph {et~al.}(2010)\citenamefont
  {Pfleiderer}, \citenamefont {Adams}, \citenamefont {Bauer}, \citenamefont
  {Biberacher}, \citenamefont {Binz}, \citenamefont {Birkelbach}, \citenamefont
  {B\"{o}ni}, \citenamefont {Franz}, \citenamefont {Georgii}, \citenamefont
  {Janoschek}, \citenamefont {Jonietz}, \citenamefont {Keller}, \citenamefont
  {Ritz}, \citenamefont {M\"uhlbauer}, \citenamefont {M\"unzer}, \citenamefont
  {Neubauer}, \citenamefont {Pedersen},\ and\ \citenamefont
  {Rosch}}]{Pfleiderer10}%
  \BibitemOpen
  \bibfield  {author} {\bibinfo {author} {\bibfnamefont {C.}~\bibnamefont
  {Pfleiderer}}, \bibinfo {author} {\bibfnamefont {T.}~\bibnamefont {Adams}},
  \bibinfo {author} {\bibfnamefont {A.}~\bibnamefont {Bauer}}, \bibinfo
  {author} {\bibfnamefont {W.}~\bibnamefont {Biberacher}}, \bibinfo {author}
  {\bibfnamefont {B.}~\bibnamefont {Binz}}, \bibinfo {author} {\bibfnamefont
  {F.}~\bibnamefont {Birkelbach}}, \bibinfo {author} {\bibfnamefont
  {P.}~\bibnamefont {B\"{o}ni}}, \bibinfo {author} {\bibfnamefont
  {C.}~\bibnamefont {Franz}}, \bibinfo {author} {\bibfnamefont
  {R.}~\bibnamefont {Georgii}}, \bibinfo {author} {\bibfnamefont
  {M.}~\bibnamefont {Janoschek}}, \bibinfo {author} {\bibfnamefont
  {F.}~\bibnamefont {Jonietz}}, \bibinfo {author} {\bibfnamefont
  {T.}~\bibnamefont {Keller}}, \bibinfo {author} {\bibfnamefont
  {R.}~\bibnamefont {Ritz}}, \bibinfo {author} {\bibfnamefont {S.}~\bibnamefont
  {M\"uhlbauer}}, \bibinfo {author} {\bibfnamefont {W.}~\bibnamefont
  {M\"unzer}}, \bibinfo {author} {\bibfnamefont {A.}~\bibnamefont {Neubauer}},
  \bibinfo {author} {\bibfnamefont {B.}~\bibnamefont {Pedersen}}, \ and\
  \bibinfo {author} {\bibfnamefont {A.}~\bibnamefont {Rosch}},\ }\href
  {http://stacks.iop.org/0953-8984/22/i=16/a=164207} {\bibfield  {journal}
  {\bibinfo  {journal} {J. Phys.: Condens. Matter}\ }\textbf {\bibinfo {volume}
  {22}},\ \bibinfo {pages} {164207} (\bibinfo {year} {2010})}\BibitemShut
  {NoStop}%
\bibitem [{\citenamefont {Yu}\ \emph {et~al.}(2010)\citenamefont {Yu},
  \citenamefont {Onose}, \citenamefont {Kanazawa}, \citenamefont {Park},
  \citenamefont {Han}, \citenamefont {Matsui}, \citenamefont {Nagaosa},\ and\
  \citenamefont {Tokura}}]{Yu2010a}%
  \BibitemOpen
  \bibfield  {author} {\bibinfo {author} {\bibfnamefont {X.~Z.}\ \bibnamefont
  {Yu}}, \bibinfo {author} {\bibfnamefont {Y.}~\bibnamefont {Onose}}, \bibinfo
  {author} {\bibfnamefont {N.}~\bibnamefont {Kanazawa}}, \bibinfo {author}
  {\bibfnamefont {J.~H.}\ \bibnamefont {Park}}, \bibinfo {author}
  {\bibfnamefont {J.~H.}\ \bibnamefont {Han}}, \bibinfo {author} {\bibfnamefont
  {Y.}~\bibnamefont {Matsui}}, \bibinfo {author} {\bibfnamefont
  {N.}~\bibnamefont {Nagaosa}}, \ and\ \bibinfo {author} {\bibfnamefont
  {Y.}~\bibnamefont {Tokura}},\ }\href {\doibase 10.1038/nature09124}
  {\bibfield  {journal} {\bibinfo  {journal} {Nature}\ }\textbf {\bibinfo
  {volume} {465}},\ \bibinfo {pages} {901} (\bibinfo {year}
  {2010})}\BibitemShut {NoStop}%
\bibitem [{\citenamefont {Yu}\ \emph {et~al.}(2011)\citenamefont {Yu},
  \citenamefont {Kanazawa}, \citenamefont {Onose}, \citenamefont {Kimoto},
  \citenamefont {Zhang}, \citenamefont {Ishiwata}, \citenamefont {Matsui},\
  and\ \citenamefont {Tokura}}]{Yu2011}%
  \BibitemOpen
  \bibfield  {author} {\bibinfo {author} {\bibfnamefont {X.~Z.}\ \bibnamefont
  {Yu}}, \bibinfo {author} {\bibfnamefont {N.}~\bibnamefont {Kanazawa}},
  \bibinfo {author} {\bibfnamefont {Y.}~\bibnamefont {Onose}}, \bibinfo
  {author} {\bibfnamefont {K.}~\bibnamefont {Kimoto}}, \bibinfo {author}
  {\bibfnamefont {W.~Z.}\ \bibnamefont {Zhang}}, \bibinfo {author}
  {\bibfnamefont {S.}~\bibnamefont {Ishiwata}}, \bibinfo {author}
  {\bibfnamefont {Y.}~\bibnamefont {Matsui}}, \ and\ \bibinfo {author}
  {\bibfnamefont {Y.}~\bibnamefont {Tokura}},\ }\href {\doibase
  10.1038/nmat2916} {\bibfield  {journal} {\bibinfo  {journal} {Nature
  Materials}\ }\textbf {\bibinfo {volume} {10}},\ \bibinfo {pages} {106}
  (\bibinfo {year} {2011})}\BibitemShut {NoStop}%
\bibitem [{\citenamefont {Heinze}\ \emph {et~al.}(2011)\citenamefont {Heinze},
  \citenamefont {Bergmann}, \citenamefont {Menzel}, \citenamefont {Brede},
  \citenamefont {Kubetzka}, \citenamefont {Wiesendanger}, \citenamefont
  {Bihlmayer},\ and\ \citenamefont {Blügel}}]{Heinze2011}%
  \BibitemOpen
  \bibfield  {author} {\bibinfo {author} {\bibfnamefont {S.}~\bibnamefont
  {Heinze}}, \bibinfo {author} {\bibfnamefont {K.~v.}\ \bibnamefont
  {Bergmann}}, \bibinfo {author} {\bibfnamefont {M.}~\bibnamefont {Menzel}},
  \bibinfo {author} {\bibfnamefont {J.}~\bibnamefont {Brede}}, \bibinfo
  {author} {\bibfnamefont {A.}~\bibnamefont {Kubetzka}}, \bibinfo {author}
  {\bibfnamefont {R.}~\bibnamefont {Wiesendanger}}, \bibinfo {author}
  {\bibfnamefont {G.}~\bibnamefont {Bihlmayer}}, \ and\ \bibinfo {author}
  {\bibfnamefont {S.}~\bibnamefont {Blügel}},\ }\href {\doibase
  10.1038/nphys2045} {\bibfield  {journal} {\bibinfo  {journal} {Nature
  Physics}\ }\textbf {\bibinfo {volume} {7}},\ \bibinfo {pages} {713} (\bibinfo
  {year} {2011})}\BibitemShut {NoStop}%
\bibitem [{\citenamefont {Seki}\ \emph
  {et~al.}(2012{\natexlab{a}})\citenamefont {Seki}, \citenamefont {Yu},
  \citenamefont {Ishiwata},\ and\ \citenamefont {Tokura}}]{Seki2012}%
  \BibitemOpen
  \bibfield  {author} {\bibinfo {author} {\bibfnamefont {S.}~\bibnamefont
  {Seki}}, \bibinfo {author} {\bibfnamefont {X.~Z.}\ \bibnamefont {Yu}},
  \bibinfo {author} {\bibfnamefont {S.}~\bibnamefont {Ishiwata}}, \ and\
  \bibinfo {author} {\bibfnamefont {Y.}~\bibnamefont {Tokura}},\ }\href
  {\doibase 10.1126/science.1214143} {\bibfield  {journal} {\bibinfo  {journal}
  {Science}\ }\textbf {\bibinfo {volume} {336}},\ \bibinfo {pages} {198}
  (\bibinfo {year} {2012}{\natexlab{a}})}\BibitemShut {NoStop}%
\bibitem [{\citenamefont {Bogdanov}\ and\ \citenamefont
  {Yablonskii}(1989)}]{Bogdanov89}%
  \BibitemOpen
  \bibfield  {author} {\bibinfo {author} {\bibfnamefont {A.~N.}\ \bibnamefont
  {Bogdanov}}\ and\ \bibinfo {author} {\bibfnamefont {D.~A.}\ \bibnamefont
  {Yablonskii}},\ }\href@noop {} {\bibfield  {journal} {\bibinfo  {journal}
  {Sov. Phys. JETP}\ }\textbf {\bibinfo {volume} {68}},\ \bibinfo {pages} {101}
  (\bibinfo {year} {1989})}\BibitemShut {NoStop}%
\bibitem [{\citenamefont {Bogdanov}\ and\ \citenamefont
  {Hubert}(1994)}]{Bogdanov94}%
  \BibitemOpen
  \bibfield  {author} {\bibinfo {author} {\bibfnamefont {A.}~\bibnamefont
  {Bogdanov}}\ and\ \bibinfo {author} {\bibfnamefont {A.}~\bibnamefont
  {Hubert}},\ }\href@noop {} {\bibfield  {journal} {\bibinfo  {journal} {J.
  Magn. Magn. Mater.}\ }\textbf {\bibinfo {volume} {138}},\ \bibinfo {pages}
  {255} (\bibinfo {year} {1994})}\BibitemShut {NoStop}%
\bibitem [{\citenamefont {R\"{o}\ss~ler}\ \emph {et~al.}(2006)\citenamefont
  {R\"{o}\ss~ler}, \citenamefont {Bogdanov},\ and\ \citenamefont
  {Pfleiderer}}]{Rosler2006}%
  \BibitemOpen
  \bibfield  {author} {\bibinfo {author} {\bibfnamefont {U.~K.}\ \bibnamefont
  {R\"{o}\ss~ler}}, \bibinfo {author} {\bibfnamefont {A.~N.}\ \bibnamefont
  {Bogdanov}}, \ and\ \bibinfo {author} {\bibfnamefont {C.}~\bibnamefont
  {Pfleiderer}},\ }\href {\doibase 10.1038/nature05056} {\bibfield  {journal}
  {\bibinfo  {journal} {Nature}\ }\textbf {\bibinfo {volume} {442}},\ \bibinfo
  {pages} {797} (\bibinfo {year} {2006})}\BibitemShut {NoStop}%
\bibitem [{\citenamefont {Yu}\ \emph {et~al.}(2012{\natexlab{a}})\citenamefont
  {Yu}, \citenamefont {Mostovoy}, \citenamefont {Tokunaga}, \citenamefont
  {Zhang}, \citenamefont {Kimoto}, \citenamefont {Matsui}, \citenamefont
  {Kaneko}, \citenamefont {Nagaosa},\ and\ \citenamefont {Tokura}}]{Yu2012b}%
  \BibitemOpen
  \bibfield  {author} {\bibinfo {author} {\bibfnamefont {X.}~\bibnamefont
  {Yu}}, \bibinfo {author} {\bibfnamefont {M.}~\bibnamefont {Mostovoy}},
  \bibinfo {author} {\bibfnamefont {Y.}~\bibnamefont {Tokunaga}}, \bibinfo
  {author} {\bibfnamefont {W.}~\bibnamefont {Zhang}}, \bibinfo {author}
  {\bibfnamefont {K.}~\bibnamefont {Kimoto}}, \bibinfo {author} {\bibfnamefont
  {Y.}~\bibnamefont {Matsui}}, \bibinfo {author} {\bibfnamefont
  {Y.}~\bibnamefont {Kaneko}}, \bibinfo {author} {\bibfnamefont
  {N.}~\bibnamefont {Nagaosa}}, \ and\ \bibinfo {author} {\bibfnamefont
  {Y.}~\bibnamefont {Tokura}},\ }\href {\doibase 10.1073/pnas.1118496109}
  {\bibfield  {journal} {\bibinfo  {journal} {{PNAS}}\ }\textbf {\bibinfo
  {volume} {109}},\ \bibinfo {pages} {8856} (\bibinfo {year}
  {2012}{\natexlab{a}})}\BibitemShut {NoStop}%
\bibitem [{\citenamefont {Finazzi}\ \emph {et~al.}(2013)\citenamefont
  {Finazzi}, \citenamefont {Savoini}, \citenamefont {Khorsand}, \citenamefont
  {Tsukamoto}, \citenamefont {Itoh}, \citenamefont {Duò}, \citenamefont
  {Kirilyuk}, \citenamefont {Rasing},\ and\ \citenamefont
  {Ezawa}}]{Finazzi2013}%
  \BibitemOpen
  \bibfield  {author} {\bibinfo {author} {\bibfnamefont {M.}~\bibnamefont
  {Finazzi}}, \bibinfo {author} {\bibfnamefont {M.}~\bibnamefont {Savoini}},
  \bibinfo {author} {\bibfnamefont {A.~R.}\ \bibnamefont {Khorsand}}, \bibinfo
  {author} {\bibfnamefont {A.}~\bibnamefont {Tsukamoto}}, \bibinfo {author}
  {\bibfnamefont {A.}~\bibnamefont {Itoh}}, \bibinfo {author} {\bibfnamefont
  {L.}~\bibnamefont {Duò}}, \bibinfo {author} {\bibfnamefont {A.}~\bibnamefont
  {Kirilyuk}}, \bibinfo {author} {\bibfnamefont {T.}~\bibnamefont {Rasing}}, \
  and\ \bibinfo {author} {\bibfnamefont {M.}~\bibnamefont {Ezawa}},\ }\href
  {\doibase 10.1103/PhysRevLett.110.177205} {\bibfield  {journal} {\bibinfo
  {journal} {Phys. Rev. Lett.}\ }\textbf {\bibinfo {volume} {110}},\ \bibinfo
  {pages} {177205} (\bibinfo {year} {2013})}\BibitemShut {NoStop}%
\bibitem [{\citenamefont {Bazaliy}\ \emph {et~al.}(1998)\citenamefont
  {Bazaliy}, \citenamefont {Jones},\ and\ \citenamefont {Zhang}}]{Bazaliy98}%
  \BibitemOpen
  \bibfield  {author} {\bibinfo {author} {\bibfnamefont {Y.~B.}\ \bibnamefont
  {Bazaliy}}, \bibinfo {author} {\bibfnamefont {B.~A.}\ \bibnamefont {Jones}},
  \ and\ \bibinfo {author} {\bibfnamefont {S.-C.}\ \bibnamefont {Zhang}},\
  }\href {\doibase 10.1103/PhysRevB.57.R3213} {\bibfield  {journal} {\bibinfo
  {journal} {Phys. Rev. B}\ }\textbf {\bibinfo {volume} {57}},\ \bibinfo
  {pages} {R3213} (\bibinfo {year} {1998})}\BibitemShut {NoStop}%
\bibitem [{\citenamefont {Li}\ and\ \citenamefont {Zhang}(2004)}]{Li04}%
  \BibitemOpen
  \bibfield  {author} {\bibinfo {author} {\bibfnamefont {Z.}~\bibnamefont
  {Li}}\ and\ \bibinfo {author} {\bibfnamefont {S.}~\bibnamefont {Zhang}},\
  }\href {\doibase 10.1103/PhysRevLett.92.207203} {\bibfield  {journal}
  {\bibinfo  {journal} {Phys. Rev. Lett.}\ }\textbf {\bibinfo {volume} {92}},\
  \bibinfo {pages} {207203} (\bibinfo {year} {2004})}\BibitemShut {NoStop}%
\bibitem [{\citenamefont {Tatara}\ \emph {et~al.}(2008)\citenamefont {Tatara},
  \citenamefont {Kohno},\ and\ \citenamefont {Shibata}}]{Tatara2008}%
  \BibitemOpen
  \bibfield  {author} {\bibinfo {author} {\bibfnamefont {G.}~\bibnamefont
  {Tatara}}, \bibinfo {author} {\bibfnamefont {H.}~\bibnamefont {Kohno}}, \
  and\ \bibinfo {author} {\bibfnamefont {J.}~\bibnamefont {Shibata}},\ }\href
  {\doibase 10.1016/j.physrep.2008.07.003} {\bibfield  {journal} {\bibinfo
  {journal} {Phys. Rep.}\ }\textbf {\bibinfo {volume} {468}},\ \bibinfo {pages}
  {213} (\bibinfo {year} {2008})}\BibitemShut {NoStop}%
\bibitem [{\citenamefont {Seki}\ \emph
  {et~al.}(2012{\natexlab{b}})\citenamefont {Seki}, \citenamefont {Ishiwata},\
  and\ \citenamefont {Tokura}}]{Seki2012b}%
  \BibitemOpen
  \bibfield  {author} {\bibinfo {author} {\bibfnamefont {S.}~\bibnamefont
  {Seki}}, \bibinfo {author} {\bibfnamefont {S.}~\bibnamefont {Ishiwata}}, \
  and\ \bibinfo {author} {\bibfnamefont {Y.}~\bibnamefont {Tokura}},\ }\href
  {\doibase 10.1103/PhysRevB.86.060403} {\bibfield  {journal} {\bibinfo
  {journal} {Phys. Rev. B}\ }\textbf {\bibinfo {volume} {86}},\ \bibinfo
  {pages} {060403} (\bibinfo {year} {2012}{\natexlab{b}})}\BibitemShut
  {NoStop}%
\bibitem [{\citenamefont {Liu}\ \emph {et~al.}(2013)\citenamefont {Liu},
  \citenamefont {Li},\ and\ \citenamefont {Han}}]{Liu2013b}%
  \BibitemOpen
  \bibfield  {author} {\bibinfo {author} {\bibfnamefont {Y.-H.}\ \bibnamefont
  {Liu}}, \bibinfo {author} {\bibfnamefont {Y.-Q.}\ \bibnamefont {Li}}, \ and\
  \bibinfo {author} {\bibfnamefont {J.~H.}\ \bibnamefont {Han}},\ }\href
  {\doibase 10.1103/PhysRevB.87.100402} {\bibfield  {journal} {\bibinfo
  {journal} {Phys. Rev. B}\ }\textbf {\bibinfo {volume} {87}},\ \bibinfo
  {pages} {100402} (\bibinfo {year} {2013})}\BibitemShut {NoStop}%
\bibitem [{\citenamefont {Jonietz}\ \emph {et~al.}(2010)\citenamefont
  {Jonietz}, \citenamefont {M\"uhlbauer}, \citenamefont {Pfleiderer},
  \citenamefont {Neubauer}, \citenamefont {M\"unzer}, \citenamefont {Bauer},
  \citenamefont {Adams}, \citenamefont {Georgii}, \citenamefont {B\"oni},
  \citenamefont {Duine}, \citenamefont {Everschor}, \citenamefont {Garst},\
  and\ \citenamefont {Rosch}}]{Jonietz2010}%
  \BibitemOpen
  \bibfield  {author} {\bibinfo {author} {\bibfnamefont {F.}~\bibnamefont
  {Jonietz}}, \bibinfo {author} {\bibfnamefont {S.}~\bibnamefont
  {M\"uhlbauer}}, \bibinfo {author} {\bibfnamefont {C.}~\bibnamefont
  {Pfleiderer}}, \bibinfo {author} {\bibfnamefont {A.}~\bibnamefont
  {Neubauer}}, \bibinfo {author} {\bibfnamefont {W.}~\bibnamefont {M\"unzer}},
  \bibinfo {author} {\bibfnamefont {A.}~\bibnamefont {Bauer}}, \bibinfo
  {author} {\bibfnamefont {T.}~\bibnamefont {Adams}}, \bibinfo {author}
  {\bibfnamefont {R.}~\bibnamefont {Georgii}}, \bibinfo {author} {\bibfnamefont
  {P.}~\bibnamefont {B\"oni}}, \bibinfo {author} {\bibfnamefont {R.~A.}\
  \bibnamefont {Duine}}, \bibinfo {author} {\bibfnamefont {K.}~\bibnamefont
  {Everschor}}, \bibinfo {author} {\bibfnamefont {M.}~\bibnamefont {Garst}}, \
  and\ \bibinfo {author} {\bibfnamefont {A.}~\bibnamefont {Rosch}},\ }\href
  {\doibase 10.1126/science.1195709} {\bibfield  {journal} {\bibinfo  {journal}
  {Science}\ }\textbf {\bibinfo {volume} {330}},\ \bibinfo {pages} {1648}
  (\bibinfo {year} {2010})}\BibitemShut {NoStop}%
\bibitem [{\citenamefont {Yu}\ \emph {et~al.}(2012{\natexlab{b}})\citenamefont
  {Yu}, \citenamefont {Kanazawa}, \citenamefont {Zhang}, \citenamefont {Nagai},
  \citenamefont {Hara}, \citenamefont {Kimoto}, \citenamefont {Matsui},
  \citenamefont {Onose},\ and\ \citenamefont {Tokura}}]{Yu2012}%
  \BibitemOpen
  \bibfield  {author} {\bibinfo {author} {\bibfnamefont {X.~Z.}\ \bibnamefont
  {Yu}}, \bibinfo {author} {\bibfnamefont {N.}~\bibnamefont {Kanazawa}},
  \bibinfo {author} {\bibfnamefont {W.~Z.}\ \bibnamefont {Zhang}}, \bibinfo
  {author} {\bibfnamefont {T.}~\bibnamefont {Nagai}}, \bibinfo {author}
  {\bibfnamefont {T.}~\bibnamefont {Hara}}, \bibinfo {author} {\bibfnamefont
  {K.}~\bibnamefont {Kimoto}}, \bibinfo {author} {\bibfnamefont
  {Y.}~\bibnamefont {Matsui}}, \bibinfo {author} {\bibfnamefont
  {Y.}~\bibnamefont {Onose}}, \ and\ \bibinfo {author} {\bibfnamefont
  {Y.}~\bibnamefont {Tokura}},\ }\href {\doibase 10.1038/ncomms1990} {\bibfield
   {journal} {\bibinfo  {journal} {Nature Communications}\ }\textbf {\bibinfo
  {volume} {3}},\ \bibinfo {pages} {988} (\bibinfo {year}
  {2012}{\natexlab{b}})}\BibitemShut {NoStop}%
\bibitem [{\citenamefont {Schulz}\ \emph {et~al.}(2012)\citenamefont {Schulz},
  \citenamefont {Ritz}, \citenamefont {Bauer}, \citenamefont {Halder},
  \citenamefont {Wagner}, \citenamefont {Franz}, \citenamefont {Pfleiderer},
  \citenamefont {Everschor}, \citenamefont {Garst},\ and\ \citenamefont
  {Rosch}}]{Schulz2012}%
  \BibitemOpen
  \bibfield  {author} {\bibinfo {author} {\bibfnamefont {T.}~\bibnamefont
  {Schulz}}, \bibinfo {author} {\bibfnamefont {R.}~\bibnamefont {Ritz}},
  \bibinfo {author} {\bibfnamefont {A.}~\bibnamefont {Bauer}}, \bibinfo
  {author} {\bibfnamefont {M.}~\bibnamefont {Halder}}, \bibinfo {author}
  {\bibfnamefont {M.}~\bibnamefont {Wagner}}, \bibinfo {author} {\bibfnamefont
  {C.}~\bibnamefont {Franz}}, \bibinfo {author} {\bibfnamefont
  {C.}~\bibnamefont {Pfleiderer}}, \bibinfo {author} {\bibfnamefont
  {K.}~\bibnamefont {Everschor}}, \bibinfo {author} {\bibfnamefont
  {M.}~\bibnamefont {Garst}}, \ and\ \bibinfo {author} {\bibfnamefont
  {A.}~\bibnamefont {Rosch}},\ }\href {\doibase 10.1038/nphys2231} {\bibfield
  {journal} {\bibinfo  {journal} {Nature Physics}\ }\textbf {\bibinfo {volume}
  {8}},\ \bibinfo {pages} {301} (\bibinfo {year} {2012})}\BibitemShut {NoStop}%
\bibitem [{\citenamefont {{S. Z. Lin \emph{et.
  al.}}}(2013)}]{szlin13skyrmion4}%
  \BibitemOpen
  \bibfield  {author} {\bibinfo {author} {\bibnamefont {{S. Z. Lin \emph{et.
  al.}}}},\ }\href@noop {} {\bibfield  {journal} {\bibinfo  {journal} {in
  preparation}\ } (\bibinfo {year} {2013})}\BibitemShut {NoStop}%
\bibitem [{\citenamefont {Lin}\ \emph {et~al.}(2013{\natexlab{a}})\citenamefont
  {Lin}, \citenamefont {Reichhardt}, \citenamefont {Batista},\ and\
  \citenamefont {Saxena}}]{szlin13skyrmion2}%
  \BibitemOpen
  \bibfield  {author} {\bibinfo {author} {\bibfnamefont {S.-Z.}\ \bibnamefont
  {Lin}}, \bibinfo {author} {\bibfnamefont {C.}~\bibnamefont {Reichhardt}},
  \bibinfo {author} {\bibfnamefont {C.~D.}\ \bibnamefont {Batista}}, \ and\
  \bibinfo {author} {\bibfnamefont {A.}~\bibnamefont {Saxena}},\ }\href
  {\doibase 10.1103/PhysRevB.87.214419} {\bibfield  {journal} {\bibinfo
  {journal} {Phys. Rev. B}\ }\textbf {\bibinfo {volume} {87}},\ \bibinfo
  {pages} {214419} (\bibinfo {year} {2013}{\natexlab{a}})}\BibitemShut
  {NoStop}%
\bibitem [{\citenamefont {Iwasaki}\ \emph {et~al.}(2013)\citenamefont
  {Iwasaki}, \citenamefont {Mochizuki},\ and\ \citenamefont
  {Nagaosa}}]{Iwasaki2013}%
  \BibitemOpen
  \bibfield  {author} {\bibinfo {author} {\bibfnamefont {J.}~\bibnamefont
  {Iwasaki}}, \bibinfo {author} {\bibfnamefont {M.}~\bibnamefont {Mochizuki}},
  \ and\ \bibinfo {author} {\bibfnamefont {N.}~\bibnamefont {Nagaosa}},\ }\href
  {\doibase 10.1038/ncomms2442} {\bibfield  {journal} {\bibinfo  {journal}
  {Nature Communications}\ }\textbf {\bibinfo {volume} {4}},\ \bibinfo {pages}
  {1463} (\bibinfo {year} {2013})}\BibitemShut {NoStop}%
\bibitem [{\citenamefont {Makhfudz}\ \emph {et~al.}(2012)\citenamefont
  {Makhfudz}, \citenamefont {Kr\"uger},\ and\ \citenamefont
  {Tchernyshyov}}]{Makhfudz2012}%
  \BibitemOpen
  \bibfield  {author} {\bibinfo {author} {\bibfnamefont {I.}~\bibnamefont
  {Makhfudz}}, \bibinfo {author} {\bibfnamefont {B.}~\bibnamefont {Kr\"uger}},
  \ and\ \bibinfo {author} {\bibfnamefont {O.}~\bibnamefont {Tchernyshyov}},\
  }\href {\doibase 10.1103/PhysRevLett.109.217201} {\bibfield  {journal}
  {\bibinfo  {journal} {Phys. Rev. Lett.}\ }\textbf {\bibinfo {volume} {109}},\
  \bibinfo {pages} {217201} (\bibinfo {year} {2012})}\BibitemShut {NoStop}%
\bibitem [{\citenamefont {Mochizuki}(2012)}]{Mochizuki2012}%
  \BibitemOpen
  \bibfield  {author} {\bibinfo {author} {\bibfnamefont {M.}~\bibnamefont
  {Mochizuki}},\ }\href {\doibase 10.1103/PhysRevLett.108.017601} {\bibfield
  {journal} {\bibinfo  {journal} {Phys. Rev. Lett.}\ }\textbf {\bibinfo
  {volume} {108}},\ \bibinfo {pages} {017601} (\bibinfo {year}
  {2012})}\BibitemShut {NoStop}%
\bibitem [{\citenamefont {Blatter}\ \emph {et~al.}(1994)\citenamefont
  {Blatter}, \citenamefont {Feigelman}, \citenamefont {Geshkenbein},
  \citenamefont {Larkin},\ and\ \citenamefont {Vinokur}}]{Blatter94}%
  \BibitemOpen
  \bibfield  {author} {\bibinfo {author} {\bibfnamefont {G.}~\bibnamefont
  {Blatter}}, \bibinfo {author} {\bibfnamefont {M.~V.}\ \bibnamefont
  {Feigelman}}, \bibinfo {author} {\bibfnamefont {V.~B.}\ \bibnamefont
  {Geshkenbein}}, \bibinfo {author} {\bibfnamefont {A.~I.}\ \bibnamefont
  {Larkin}}, \ and\ \bibinfo {author} {\bibfnamefont {V.~M.}\ \bibnamefont
  {Vinokur}},\ }\href@noop {} {\bibfield  {journal} {\bibinfo  {journal} {Rev.
  Mod. Phys.}\ }\textbf {\bibinfo {volume} {66}},\ \bibinfo {pages} {1125}
  (\bibinfo {year} {1994})}\BibitemShut {NoStop}%
\bibitem [{\citenamefont {Kopnin}\ and\ \citenamefont
  {Lopatin}(1995)}]{Kopnin1995}%
  \BibitemOpen
  \bibfield  {author} {\bibinfo {author} {\bibfnamefont {N.~B.}\ \bibnamefont
  {Kopnin}}\ and\ \bibinfo {author} {\bibfnamefont {A.~V.}\ \bibnamefont
  {Lopatin}},\ }\href {\doibase 10.1103/PhysRevB.51.15291} {\bibfield
  {journal} {\bibinfo  {journal} {Phys. Rev. B}\ }\textbf {\bibinfo {volume}
  {51}},\ \bibinfo {pages} {15291} (\bibinfo {year} {1995})}\BibitemShut
  {NoStop}%
\bibitem [{\citenamefont {Bulaevskii}\ \emph {et~al.}(1995)\citenamefont
  {Bulaevskii}, \citenamefont {Larkin}, \citenamefont {Maley},\ and\
  \citenamefont {Vinokur}}]{Bulaevskii1995}%
  \BibitemOpen
  \bibfield  {author} {\bibinfo {author} {\bibfnamefont {L.~N.}\ \bibnamefont
  {Bulaevskii}}, \bibinfo {author} {\bibfnamefont {A.~I.}\ \bibnamefont
  {Larkin}}, \bibinfo {author} {\bibfnamefont {M.~P.}\ \bibnamefont {Maley}}, \
  and\ \bibinfo {author} {\bibfnamefont {V.~M.}\ \bibnamefont {Vinokur}},\
  }\href {\doibase 10.1103/PhysRevB.52.9205} {\bibfield  {journal} {\bibinfo
  {journal} {Phys. Rev. B}\ }\textbf {\bibinfo {volume} {52}},\ \bibinfo
  {pages} {9205} (\bibinfo {year} {1995})}\BibitemShut {NoStop}%
\bibitem [{\citenamefont {Zang}\ \emph {et~al.}(2011)\citenamefont {Zang},
  \citenamefont {Mostovoy}, \citenamefont {Han},\ and\ \citenamefont
  {Nagaosa}}]{Zang11}%
  \BibitemOpen
  \bibfield  {author} {\bibinfo {author} {\bibfnamefont {J.}~\bibnamefont
  {Zang}}, \bibinfo {author} {\bibfnamefont {M.}~\bibnamefont {Mostovoy}},
  \bibinfo {author} {\bibfnamefont {J.~H.}\ \bibnamefont {Han}}, \ and\
  \bibinfo {author} {\bibfnamefont {N.}~\bibnamefont {Nagaosa}},\ }\href
  {\doibase 10.1103/PhysRevLett.107.136804} {\bibfield  {journal} {\bibinfo
  {journal} {Phys. Rev. Lett.}\ }\textbf {\bibinfo {volume} {107}},\ \bibinfo
  {pages} {136804} (\bibinfo {year} {2011})}\BibitemShut {NoStop}%
\bibitem [{\citenamefont {Civale}\ \emph {et~al.}(1991)\citenamefont {Civale},
  \citenamefont {Marwick}, \citenamefont {Worthington}, \citenamefont {Kirk},
  \citenamefont {Thompson}, \citenamefont {Krusin-Elbaum}, \citenamefont {Sun},
  \citenamefont {Clem},\ and\ \citenamefont {Holtzberg}}]{Civale1991}%
  \BibitemOpen
  \bibfield  {author} {\bibinfo {author} {\bibfnamefont {L.}~\bibnamefont
  {Civale}}, \bibinfo {author} {\bibfnamefont {A.~D.}\ \bibnamefont {Marwick}},
  \bibinfo {author} {\bibfnamefont {T.~K.}\ \bibnamefont {Worthington}},
  \bibinfo {author} {\bibfnamefont {M.~A.}\ \bibnamefont {Kirk}}, \bibinfo
  {author} {\bibfnamefont {J.~R.}\ \bibnamefont {Thompson}}, \bibinfo {author}
  {\bibfnamefont {L.}~\bibnamefont {Krusin-Elbaum}}, \bibinfo {author}
  {\bibfnamefont {Y.}~\bibnamefont {Sun}}, \bibinfo {author} {\bibfnamefont
  {J.~R.}\ \bibnamefont {Clem}}, \ and\ \bibinfo {author} {\bibfnamefont
  {F.}~\bibnamefont {Holtzberg}},\ }\href {\doibase 10.1103/PhysRevLett.67.648}
  {\bibfield  {journal} {\bibinfo  {journal} {Phys. Rev. Lett.}\ }\textbf
  {\bibinfo {volume} {67}},\ \bibinfo {pages} {648} (\bibinfo {year}
  {1991})}\BibitemShut {NoStop}%
\bibitem [{\citenamefont {Konczykowski}\ \emph {et~al.}(1991)\citenamefont
  {Konczykowski}, \citenamefont {Rullier-Albenque}, \citenamefont {Yacoby},
  \citenamefont {Shaulov}, \citenamefont {Yeshurun},\ and\ \citenamefont
  {Lejay}}]{Konczykowski1991}%
  \BibitemOpen
  \bibfield  {author} {\bibinfo {author} {\bibfnamefont {M.}~\bibnamefont
  {Konczykowski}}, \bibinfo {author} {\bibfnamefont {F.}~\bibnamefont
  {Rullier-Albenque}}, \bibinfo {author} {\bibfnamefont {E.~R.}\ \bibnamefont
  {Yacoby}}, \bibinfo {author} {\bibfnamefont {A.}~\bibnamefont {Shaulov}},
  \bibinfo {author} {\bibfnamefont {Y.}~\bibnamefont {Yeshurun}}, \ and\
  \bibinfo {author} {\bibfnamefont {P.}~\bibnamefont {Lejay}},\ }\href
  {\doibase 10.1103/PhysRevB.44.7167} {\bibfield  {journal} {\bibinfo
  {journal} {Phys. Rev. B}\ }\textbf {\bibinfo {volume} {44}},\ \bibinfo
  {pages} {7167} (\bibinfo {year} {1991})}\BibitemShut {NoStop}%
\bibitem [{\citenamefont {Feigel'man}\ \emph {et~al.}(1993)\citenamefont
  {Feigel'man}, \citenamefont {Geshkenbein}, \citenamefont {Larkin},\ and\
  \citenamefont {Levit}}]{Feigelman1993}%
  \BibitemOpen
  \bibfield  {author} {\bibinfo {author} {\bibfnamefont {M.~V.}\ \bibnamefont
  {Feigel'man}}, \bibinfo {author} {\bibfnamefont {V.~B.}\ \bibnamefont
  {Geshkenbein}}, \bibinfo {author} {\bibfnamefont {A.~I.}\ \bibnamefont
  {Larkin}}, \ and\ \bibinfo {author} {\bibfnamefont {S.}~\bibnamefont
  {Levit}},\ }\href@noop {} {\bibfield  {journal} {\bibinfo  {journal} {JETP
  Lett.}\ }\textbf {\bibinfo {volume} {57}},\ \bibinfo {pages} {711} (\bibinfo
  {year} {1993})}\BibitemShut {NoStop}%
\bibitem [{\citenamefont {Lin}\ \emph {et~al.}(2013{\natexlab{b}})\citenamefont
  {Lin}, \citenamefont {Reichhardt},\ and\ \citenamefont
  {Saxena}}]{szlin13skyrmion3}%
  \BibitemOpen
  \bibfield  {author} {\bibinfo {author} {\bibfnamefont {S.-Z.}\ \bibnamefont
  {Lin}}, \bibinfo {author} {\bibfnamefont {C.}~\bibnamefont {Reichhardt}}, \
  and\ \bibinfo {author} {\bibfnamefont {A.}~\bibnamefont {Saxena}},\ }\href
  {\doibase doi:10.1063/1.4809751} {\bibfield  {journal} {\bibinfo  {journal}
  {Appl. Phys. Lett.}\ }\textbf {\bibinfo {volume} {102}},\ \bibinfo {pages}
  {222405} (\bibinfo {year} {2013}{\natexlab{b}})}\BibitemShut {NoStop}%
\bibitem [{\citenamefont {Han}\ \emph {et~al.}(2010)\citenamefont {Han},
  \citenamefont {Zang}, \citenamefont {Yang}, \citenamefont {Park},\ and\
  \citenamefont {Nagaosa}}]{Han10}%
  \BibitemOpen
  \bibfield  {author} {\bibinfo {author} {\bibfnamefont {J.~H.}\ \bibnamefont
  {Han}}, \bibinfo {author} {\bibfnamefont {J.}~\bibnamefont {Zang}}, \bibinfo
  {author} {\bibfnamefont {Z.}~\bibnamefont {Yang}}, \bibinfo {author}
  {\bibfnamefont {J.-H.}\ \bibnamefont {Park}}, \ and\ \bibinfo {author}
  {\bibfnamefont {N.}~\bibnamefont {Nagaosa}},\ }\href {\doibase
  10.1103/PhysRevB.82.094429} {\bibfield  {journal} {\bibinfo  {journal} {Phys.
  Rev. B}\ }\textbf {\bibinfo {volume} {82}},\ \bibinfo {pages} {094429}
  (\bibinfo {year} {2010})}\BibitemShut {NoStop}%
\bibitem [{\citenamefont {R\"{o}\ss~ler}\ \emph {et~al.}(2011)\citenamefont
  {R\"{o}\ss~ler}, \citenamefont {Leonov},\ and\ \citenamefont
  {Bogdanov}}]{Rossler2011}%
  \BibitemOpen
  \bibfield  {author} {\bibinfo {author} {\bibfnamefont {U.~K.}\ \bibnamefont
  {R\"{o}\ss~ler}}, \bibinfo {author} {\bibfnamefont {A.~A.}\ \bibnamefont
  {Leonov}}, \ and\ \bibinfo {author} {\bibfnamefont {A.~N.}\ \bibnamefont
  {Bogdanov}},\ }\href {http://stacks.iop.org/1742-6596/303/i=1/a=012105}
  {\bibfield  {journal} {\bibinfo  {journal} {J. Phys.: Conference Series}\
  }\textbf {\bibinfo {volume} {303}},\ \bibinfo {pages} {012105} (\bibinfo
  {year} {2011})}\BibitemShut {NoStop}%
\end{thebibliography}
%

\end{document}